\documentclass[preprint2]{aastex}
\usepackage{natbib}
\begin{document}

\title{\large{\rm{AN ECLECTIC VIEW OF OUR MILKY WAY GALAXY}}}
\author{\small David G. Turner$^1$}
\affil{$^1${\footnotesize Saint Mary's University, Halifax, Nova Scotia, Canada.}}
\email{turner@ap.smu.ca}
\begin{abstract}
The nature of our Milky Way Galaxy is reexamined from an eclectic point of view. Evidence for a central bar, for example, is not reflected in the distribution of RR Lyrae variables in the central bulge [4,5], and it is not clear if either a 2-armed or 4-armed spiral pattern is appropriate for the spiral arms. Radial velocity mapping of the Galaxy using radio H I, H II, or CO observations is compromised by the assumptions adopted for simple Galactic rotation. The Sun's local standard of rest (LSR) velocity is $\sim 14$ km s$^{-1}$ rather than 20 km s$^{-1}$, the local circular velocity is $251 \pm 9$ km s$^{-1}$ rather than 220 km s$^{-1}$, and young groups of stars exhibit a 10--20 km s$^{-1}$ ``kick'' relative to what is expected from Galactic rotation. By implication, the same may be true for star-forming gas clouds affected by the Galaxy's spiral density wave, raising concerns about their use for mapping spiral arms. Proper motion data in conjunction with the newly-estimated velocity components for the Sun's motion imply a distance to the Galactic centre of $R_0=8.34\pm0.27$ kpc , consistent with recent estimates which average $8.24\pm0.09$ kpc. A cosinusoidal Galactic potential is not ruled out by observations of open star clusters. The planetary nebula cluster Bica 6, for example, has a near-escape orbit for a Newtonian potential, but a near-normal orbit in a cosinusoidal potential field. The nearby cluster Collinder 464 also displays unusually large tidal effects consistent with those expected for a cosinusoidal potential. A standard Newtonian version of the Virial Theorem for star clusters yields very reasonable masses ($\sim 3 \times 10^{11}M_{\odot}$ and $\sim 4 \times 10^{11}M_{\odot}$) for the Milky Way and M31 subgroups of the Local Group, respectively. A cosinusoidal relation should yield identical results. 
\end{abstract}

\section{{\rm \footnotesize INTRODUCTION}}
Introductory astronomy textbooks usually contain artist impressions of what our Galaxy looks like as viewed by an observer located well above its central plane (e.g., Wikipedia entry for Milky Way). Such images vary from source to source, but normally picture the Milky Way as a two-armed or four-armed spiral with a prominent central bar inclined by $\sim 30^{\circ}$ from the line of sight towards the Galactic centre. Yet the complete delineation of the Galaxy's spiral arms remains a challenge [1,2], despite the many years that have elapsed since the early evidence from the distribution of OB-type stars presented by Morgan [3]. A mapping of RR Lyrae variables in the direction of the Galactic bulge [4] displays no evidence of a bar, for example, at the same time generating a very reasonable estimate of $8.1 \pm0.6$ kpc for the distance to the Galactic centre, $R_0$. Independent support for that conclusion has recently been provided [5], with a similar estimate of $R_0=8.33\pm0.14$. 

No hint of a central bar is seen in Sergei Gaposhkin's fanciful sketches of the Milky Way from Australia [6]. Arguments for the presence of a bar at the Galactic centre mostly postdate the 1964 introduction of the density wave model for spiral structure by Lin \& Shu [7], and its existence appears to be generally accepted (e.g., [8,9,10]). The question remains whether or not the actual picture is like the four-armed barred spirals depicted by Russeil [11] and Vall\'{e}e [12].

All arguments about the exact nature of the Milky Way as a galaxy assume that its gravitational potential is described in standard Newtonian fashion, specifically the general relativistic model of Einstein. However, a more recent development linking the physical formulations for electromagnetism and gravitation has been developed [13] that has interesting consequences if extended to the Galaxy. A cosinusoidal gravitational potential implies a different distribution of mass in the Galaxy than is the case for a strictly Newtonian potential [13,14,15,16,17], and it is of interest to explore potential observational clues that can test the hypothesis. Recent arguments for missing mass and dark matter haloes in galaxies are directly affected by the manner in which mass in the Galaxy is linked to its gravitational potential field. Do observations of the various components of the Galaxy provide any information pertinent to the question?

\section{{\rm \footnotesize SPIRAL ARM MAPPING}}
The Galaxy's spiral arms have been mapped in two independent ways using objects projected on the Galactic plane: (i) young stellar objects (supergiants, long period Cepheids, OB stars) and young open clusters, and (ii) density peaks for clouds of neutral (H I) and ionized (H II) hydrogen, as well as from the mapping of giant molecular clouds (sites of star formation) using CO. The former relies on the open cluster and Cepheid distance scales, which appear to be in good agreement [18], although there is a worrisome dependence on corrections for interstellar extinction at optical wavelengths, which can cause systematic effects. A case in point is the anomalous reddening towards Carina, described by a ratio of total-to-selective extinction of $R=A_V/E_{B-V}\simeq4$ [19,20], which makes objects in that direction appear more distant than they are. Arguments have also been made for a value of $R\simeq2.5$ towards the Galactic bulge [21]. The latter technique relies mainly on inferring distances to hydrogen and molecular clouds from their radial velocities relative to the Local Standard of Rest (LSR), in conjunction with a simple model for Galactic rotation.

\begin{figure}[!t]
\center
\includegraphics[width=0.45\textwidth]{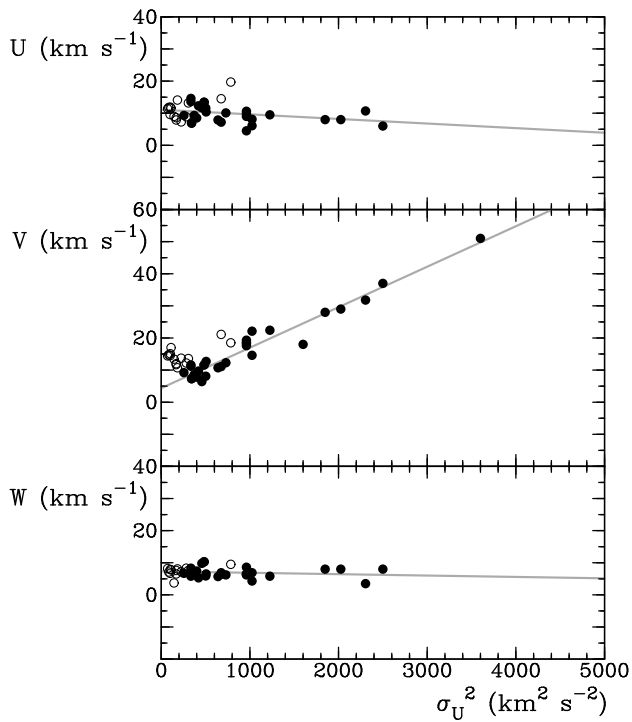}
\caption{\small{Motion of the Sun relative to different kinematic groups as a function of {\it U-}variance, where older, well-mixed groups with evolutionary ages in excess of $\sim 5 \times 10^8$ years are denoted by filled circles, younger and less well-mixed groups by open circles. Solutions for {\it U} and {\it W} incorporated data for both types of groups, that for {\it V} only older groups.}}
\label{fig1}
\end{figure}

A source of current problems with our knowledge of Galactic rotation is evident from a 1986 review for the International Astronomical Union (IAU) [22], in which the LSR velocity for the Sun is assumed to be the standard solar motion. The Newtonian version of Galactic dynamics approximates roseate stellar orbits about the Galactic centre by epicycles, in which the epicycle frequency $\kappa$ is typically greater than the orbital frequency $\omega$ [23,24]. Elliptical orbits for which $\kappa = 2\omega$ do not match the observed motions of old disk stars, which are best described by $\kappa \simeq 1.4\;\omega$. Such an approximation is not restrictive and accommodates a flat rotation curve and the observed velocity ellipsoid for old disk stars. The latter describes the peculiar motions of stars in the Galactic plane, which are most ``ellipsoidal'' for old disk stars. A distinct vertex deviation, a tilt of the velocity ellipsoid from the direction towards the Galactic centre, is observed for young stars, and is a matter of active debate.

Kinematic solutions tied to proper motion and/or radial velocity measures have been made for the Sun's motion relative to various groups of stars [23,25,26]. The standard solar motion is defined by the Sun's motion relative to the majority of catalogued stars, and amounts to $\sim19.5$ km s$^{-1}$. The basic solar motion is defined by the most frequently occurring velocity components of the Sun's motion, and amounts to $\sim15.4$ km s$^{-1}$. Although the standard solar motion has for many years been accepted as the motion of the LSR for velocity mapping of the Galaxy in H I [e.g., 27,28,29], neither it nor the basic solar motion is related to the solar motion relative to the dynamical standard of rest, which is much smaller, in the range $11-13$ km s$^{-1}$ [30,31]. Although the Mayor [30] and Oblak [31] results are frequently overlooked, there is some support for the smaller velocity from recent results for the Sun's motion relative to the interstellar medium [32], which amounts to only $\sim 2$ km s$^{-1}$ in the direction of Galactic orbital motion.

A new solution for the Sun's motion relative to the dynamical LSR was made many years ago by the author, and made use of the standard epicyclic approximation used to describe the drift of any kinematic group relative to the LSR. That should depend upon the velocity dispersion of the group in the direction towards the Galactic centre, i.e.~$\sigma_U$. The equations of Galactic dynamics [23,30] predict a drift between the kinematical and dynamical LSR that depends directly upon the square of the velocity dispersion of the group relative to the direction to the Galactic centre, namely $V_{\rm dyn} - V_{\rm kin} \propto \sigma_U^2$. An extrapolation to zero dispersion of the trend in the observed differences relative to the squares of their {\it U-}component velocity dispersions should therefore yield the Sun's motion relative to the dynamical LSR. Such analyses in earlier years always produced a dependence that led to solutions for the {\it V-}component of the Sun's motion of $\sim12$ km s$^{-1}$, so that the overall solution for the Sun's motion was $\sim19.5$ km s$^{-1}$, close to more recent solutions for the standard solar motion [see 22]. That was the conclusion reached in the Kerr \& Lynden-Bell review [22].

A closer look at the data available from [25,26] and [33] suggests a different solution, as displayed in Fig. 1. As noted by Mayor [30] and Oblak [31], young stellar groups with a small {\it U-}component dispersion display an offset from the trend displayed by older stellar groups with ages in excess of $\sim5 \times 10^8$ yr (2 Galactic orbits). Such an offset is predicted by density wave models of the Galaxy [34], and also explains the increase in the vertex deviation with decreasing age for stellar groups [30,31]. A solution from Fig. 1 that incorporates groups of all ages would indeed generate a solution for $V_{\odot}$ of order 15 km s$^{-1}$, consistent with the conclusions of Kerr \& Lynden-Bell [22]. However, a solution for $\sigma_U^2 = 0$ implied by older, well-mixed kinematic groups of Galactic stars is $V_{\odot} - V_0 \simeq +4$ km s$^{-1}$. A formal solution, i.e., for $\sigma_U^2=0$ using linear fits to the data by combining least squares and non-parametric techniques (Fig. 1), for the solar motion tied to kinematic groups of stars with ages in excess of $5 \times 10^8$ yr is:
\begin{eqnarray}
U_{\odot} = 11.1 \pm0.5\;{\rm km\;s}^{-1} \nonumber \\
V_{\odot} = +4.4 \pm0.6\;{\rm km\;s}^{-1} \nonumber \\
W_{\odot} = +7.3 \pm0.2\;{\rm km\;s}^{-1} \nonumber
\end{eqnarray}
giving a solar motion of $14.0$ km s$^{-1}$ towards Galactic coordinates $\ell_{\rm LSR} = 21^{\circ}.5$, $b_{\rm LSR} = +31^{\circ}.6$.

The solution is very similar to previous results by Mayor [30] and Oblak [31], as well as to the LSR solution for neighbouring stars found from Hipparcos proper motions [35], as summarized in Table 1. The residual motions for young, recently-created stars (less than 2 orbits) exhibit a ``kick'' of $\sim 10-20$ km s$^{-1}$ in the direction opposite Galactic rotation, likely associated with the mechanism of their creation, involving an interaction of the parent cloud of gas and dust with a spiral density wave [34]. Such dynamical effects also help to explain the streaming motions observed along the edges of prominent spiral arms [e.g., 34,36]. But earlier maps of Galactic spiral structure derived from the radial velocities of hydrogen clouds may be inherently biased because they are tied to invalid corrections for the LSR velocity of the Sun.

\begin{table}[!t]
\caption{Motion of Sun Relative to LSR.}
\vspace{0.03cm}
\center
\begin{tabular}{lccc}
\hline
\hline
$U_{\odot}$ (km/s)  & $V_{\odot}$ (km/s) & $W_{\odot}$ (km/s) & Source \\
\hline
$10.3 \pm1.0$ &$6.3\pm0.9$ &$5.9\pm0.4$ &[30] \\
$8.2 \pm1.8$ &$5.0\pm0.7$ &$5.5\pm0.4$ &[31] \\
$9.7 \pm0.3$ &$5.2\pm1.0$ &$6.7\pm0.2$ &[35] \\
$11.1 \pm0.5$ &$4.4\pm0.6$ &$7.3\pm0.2$ &This paper \\
\hline
\end{tabular}
\label{tab1}
\end{table}

A second source of potential error is the adopted value for the local circular velocity, denoting the orbital velocity expected for the dynamical LSR. Typical values for that in recent years have tended to cluster near 220 km s$^{-1}$, although such a small value cannot be reconciled with the observational evidence.

There are a variety of methods employed to establish $\theta_0$, the orbital speed of the LSR, but many depend upon a few critical assumptions that can affect the results [see 22]. Two independent methods are available for establishing $\theta_0$: by examining the motions of nearby galaxies relative to the Sun or LSR [22], and by searching for a gap in the distribution of solar motions for high-velocity stars that is expected to be the signature of ``zero-velocity'' or plunging disk stars [37]. The latter method was used by Carlberg \& Innanen to derive a solar motion of $250\pm15$ km s$^{-1}$ with respect to the gap [37]. They corrected that to $\theta_0=235\pm10$ km s$^{-1}$ by adopting $V_{\odot}=15$ km s$^{-1}$, but as noted above the true solar motion relative to the LSR is smaller. For the present value of $V_{\odot}$ and no adjustment of the uncertainty, the local circular velocity relative to plunging disk stars is $\theta_0=246\pm15$ km s$^{-1}$.

A straightforward measurement of the Sun's motion with respect to Local Group galaxies leads to implausibly large values of $\sim 294$ km s$^{-1}$ [38] or larger, as noted by Kerr \& Lynden-Bell [22]. That is because many of the neighbouring galaxies located roughly in the direction of Galactic rotation are associated with the Andromeda galaxy, M31, which tends to unduly influence the solution for $\theta_0$ because of its very large mass. Many of the galaxies spatially near M31 appear to be dynamically affected by its presence and motion. Arp [39] managed to circumvent the problem by correcting velocities of more massive galaxies in the Local Group for redshift quantization [40,41], bringing the question of the local circular velocity of the Galaxy into the controversial arena of quasar redshifts. Arp's solution of 251 km s$^{-1}$ for the Sun's motion relative to neighbouring galaxies [39] agrees closely with the Carlberg \& Innanen result, but the methodology raises questions about its validity.

It turns out there is no need to adjust the velocities of Local Group galaxies for a solution to the problem, but it is still necessary to remove M31 and its neighbouring galaxies from the solution. In the time since those earlier studies, many more faint galaxies with measured radial velocities have been added to the complement of the Local Group, and they yield a solution directly. For the present study an updated list of Local Group members and their properties was supplied by Ian Steer of the NASA/IPAC Extragalactic Database (NED). The radial velocities of sample galaxies were corrected for the Sun's motion relative to the LSR and plotted in Fig. 2 as a function of $\cos \lambda$, the cosine of the angular separation of the galaxy from the direction $\ell=90^{\circ}$, $b=0^{\circ}$.

An initial solution was obtained using least squares and non-parametric fitting of a linear relation to the data restricted to galaxies within $\sim 0.4$ Mpc, in other words galaxies close to the Milky Way and not associated with M31. That gave a value of $\theta_0=259\pm27$ km s$^{-1}$. It was noted that there were many other galaxies in more extended parts of the Local Group that followed the trend indicated by the inner group. A separate solution including those galaxies resulted in $\theta_0=252\pm17$ km s$^{-1}$. Both sets of galaxies exhibit a velocity dispersion of $\pm71$ km s$^{-1}$. The group of galaxies associated with M31 deviates from the trend for nearby galaxies by $\sim 100$ km s$^{-1}$, as do galaxies in more distant portions of the nominal Local Group, although in the opposite sense. The latter may be associated with the Hubble flow, while the former are clearly influenced by the gravitational influence of M31.

An independent solution for the local circular velocity was derived by Reid et al. [42] using the positions, parallaxes, and proper motions of Galactic radio masers in conjunction with a model for the Galaxy. Their value of $\theta_0=254\pm16$ km s$^{-1}$ agrees closely with the present results. A weighted mean of the various independent solutions (plunging disk stars, motions of nearby galaxies unaffected by M31, radio masers and Galaxy model) produces a value of $\theta_0=251\pm9$ km s$^{-1}$ as a final result.

\begin{figure}[!t]
\center
\includegraphics[width=0.45\textwidth]{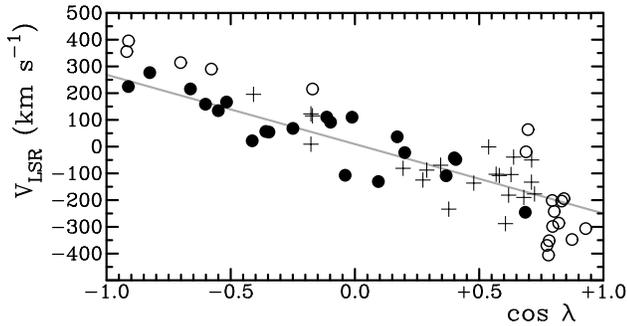}
\caption{\small{LSR-corrected velocities for galaxies of the extended Local Group plotted vs.~cosine of angular distance $\lambda$ from $\ell=90^{\circ}$, $b=0^{\circ}$. Galaxies within 400 kpc (filled circles) and other galaxies sharing their velocity trend (plus signs) produced the solution shown, while galaxies in the M31 group (far right, lower) and others at large distances (open circles) were ignored. The corresponding LSR orbital velocity is 259 km s$^{-1}$.}}
\label{fig2}
\end{figure}

A good test of the validity of this solution is to establish the distance of the Sun from the Galactic centre by comparing the inferred speeds of the Sun in the direction of Galactic rotation and perpendicular to the plane with corresponding observations of the proper motion of Sagittarius A* in $\ell$ and $b$. The two are related by the standard formula $v_t=4.74\;\mu d$, where $\mu$ is proper motion, $d$ is distance, and $v_t$ is tangential velocity. With $\mu_{\ell}=-6.379\pm0.026$ mas yr$^{-1}$ and $\mu_b=-0.202\pm0.019$ mas yr$^{-1}$ [43], the corresponding values of $R_0$ are $8.45\pm0.29$ kpc and $7.63\pm0.76$ kpc, respectively. The weighted mean of the two values is $R_0=8.34\pm0.27$ kpc.

\begin{table}[!t]
\caption{Recent Estimates for $R_0$.}
\vspace{0.03cm}
\center
\begin{tabular}{lll}
\hline
\hline
$R_0$ (kpc) & Method & Source \\
\hline
$7.94 \pm0.42$ &Sgr A* Orbits &[44] \\
$7.62 \pm0.32$ &Sgr A* Orbits &[45] \\
$8.0 \pm0.6$ &Sgr A* Orbit &[46] \\
$8.4 \pm0.4$ &Sgr A* Astrometry &[47] \\
$8.4 \pm0.6$ &Radio Masers \&  &[42] \\
& Galaxy Model & \\
$8.7 \pm0.5$ &Galaxy Model &[48] \\
$8.28 \pm0.29$ &Sgr A* Orbit &[49] \\
$7.9 \pm0.7$ &Sgr B2 Parallax &[50] \\
$8.1 \pm0.6$ &Bulge RR Lyraes &[51] \\
$8.33 \pm0.14$ &Bulge RR Lyraes &[5] \\
$8.34 \pm0.27$ &Solar Motion &This paper \\
\hline
\end{tabular}
\label{tab2}
\end{table}

A compilation of other recent estimates for $R_0$ is given in Table 2. All are consistent with a distance to the Galactic centre close to 8 kpc, although the recent trend is towards values slightly larger. A weighted mean of all Table 2 estimates is $R_0=8.24\pm0.09$ kpc, which can probably be adopted as the best estimate presently available. It is very similar to the value of 8.2 kpc advocated by many astronomers in recent years.

The point to be made, however, is that radial velocity mapping of Galactic H I, H II, and CO clouds is certain to generate biased results for Galactic spiral structure for two reasons. First, the constants for Galactic rotation adopted by the IAU in 1985 [see 22] are incorrect, being too small for $\theta_0$ (220 km s$^{-1}$ vs.~251 km s$^{-1}$), too large for the solar motion relative to the LSR (20 km s$^{-1}$ vs.~14 km s$^{-1}$), and slightly large for $R_0$ (8.5 kpc vs.~8.24 kpc). Second, the implied offset or ``kick'' from standard Galactic rotation expected [34] and observed for young objects associated with a spiral density wave implies that a simple model for Galactic rotation is insufficient for correlating velocity with distance for the Galaxy's gas clouds. The main basis for spiral arm mapping is therefore through distances to young objects observed optically, where other systematic effects can be important. 

\section{{\rm \footnotesize STAR CLUSTERS AND THE GALAXY}}
The difference between the Galaxy's gravitational potential for standard Newtonian dynamics and a cosinusoidal potential is significant enough to produce observable effects. Star clusters appear to be one means of testing such differences. A good case in point is the cluster Bica 6, at $\ell=167^{\circ}$, which has an observed radial velocity of $\sim 57$ km s$^{-1}$ [52], $\sim 56$ km s$^{-1}$ larger than expected for Galactic rotation with a flat rotation curve (Fig. 3). For a Newtonian Galactic gravitational potential that implies an orbital velocity for the cluster and its associated planetary nebula of $\sim 334$ km s$^{-1}$, very close to escape velocity. It is then a mystery how the cluster obtained such a large boost in its orbital speed relative to other open clusters in this direction. In a cosinusoidal potential, however, it is possible to consider the cluster's radial velocity as mainly the result of an excess velocity of $\sim 58$ km s$^{-1}$ in a direction away from the Galactic centre, a bit large, perhaps, but not inordinately so, for the cluster's oscillatory motion away from the 400 pc turning points [14,53]. In this scheme, Bica 6 lies less than 0.15 kpc beyond such a turning point.

\begin{figure}[!t]
\center
\includegraphics[width=0.45\textwidth]{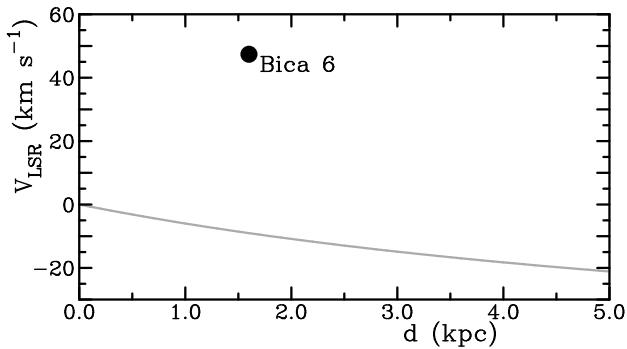}
\caption{\small{The observed radial velocity of 57 km s$^{-1}$ for Bica 6 is $\sim 56$ km s$^{-1}$ larger than expected for Galactic rotation with a flat rotation curve (gray curve).}}
\label{fig3}
\end{figure}

A Galactic cosinusoidal potential produces much stronger tidal effects than does a Newtonian potential [54]. Such effects are particularly relevant to clusters lying above or below the Galactic plane by less than 400 pc. An interesting example is the cluster Collinder 464, shown in Fig. 4. Collinder 464 is little studied, mainly because of a high declination and the fact that Collinder made a small typo in his coordinates for the cluster [55]. It is actually a degree further north than the value originally published by him. The cluster is little reddened, if at all ($E_{B-V}\simeq0.01$), and only $\sim 130$ pc distant. Its location places it just above the Galactic plane, but what is most striking is the elongation towards the Galactic plane displayed by its member stars, which comprise nearly all of the objects lying within the ellipse plotted in Fig. 4. Strongly tidally distorted star clusters are rare in the Galaxy, but Collinder 464 appears to be a good case.

\begin{figure}[!t]
\center
\includegraphics[width=0.45\textwidth]{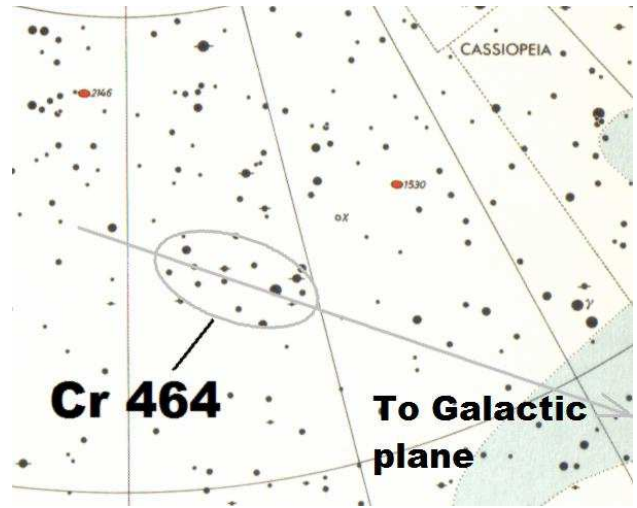}
\caption{\small{The nearby cluster Collinder 464 lies just north of the Galactic plane by $\sim 45$ pc, and is unusual in displaying a marked tidal distortion towards the Galactic plane as illustrated. Figure adapted from {\it Sky Atlas 2000.0}, first edition [56]. Used with permission from the author.}}
\label{fig4}
\end{figure}

For many years, astronomers have used a Newtonian version of the Virial Theorem suitable for Galactic objects to estimate masses for star clusters. The relevant formula is:
\[
M/M_{\odot} = \langle R({\rm pc})\rangle \left \{ \frac{\langle v^2({\rm km/s})\rangle^{\frac{1}{2}}} {4.637\times10^{-2}}\right \} ^2
\]
\noindent
where $\langle v^2 \rangle ^{\frac{1}{2}}$ is the velocity dispersion of the group in km s$^{-1}$ and $\langle R \rangle$ is the average radius of the group in parsecs (pc). The Milky Way subgroup of the Local Group used in Fig. 2 has a velocity dispersion of $\pm 71$ km s$^{-1}$ and an average radial extent of 123,000 pc, yielding $M\simeq 3 \times 10^{11}M_{\odot}$, a reasonable result for a group containing the Milky Way and and all of the other galaxies in the Milky Way subgroup of the Local Group. The M31 subgroup is more difficult to analyze. The group appears to have a velocity dispersion of $\pm 125$ km s$^{-1}$ and an average radius of 50,000 pc, yielding $M\simeq 4 \times 10^{11}M_{\odot}$, implying an overall larger mass for the galaxies in the M31 subgroup. Such a result seems reasonable.

The difference between a Newtonian gravitational potential and a cosinusoidal potential is the cosine term with its putative universal ``wavelength'' of 400 pc [53]. Since the average radius of most star clusters of $\sim 2$ pc is much smaller than 400 pc, the formula cited above should be unchanged in a cosinusoidal potential field. Likewise, the typical radius of $\sim 50,000$ pc for a small cluster of galaxies estimated above is much larger than 400 pc, so the formula should be equally suitable on galaxy cluster scales. 
The masses estimated here for the Milky Way and M31 subgroups of the Local Group, taken in isolation, are therefore unable to discriminate between cosinusoidal and Newtonian gravitational potential relations.
 
\subsection*{Acknowledgements}
\scriptsize{David Bartlett is responsible for the author's exposure to the wonders of a cosinusoidal potential in the context of the Galaxy's gravitational field. Many of the results in this study were presented at a meeting of the Canadian Astronomical Society in 1996 and at the Boston 2010 meeting of the AAS Division of Dynamical Astronomy. They were reworked for presentation at the Einstein versus Schwinger miniconference held at the AAS Anchorage meeting in June 2012. Ian Steer kindly provided up-to-date data for the analysis of $\theta_0$. Figure 4 was adapted from {\it Sky Atlas 2000.0} with permission from Wil Tirion.}

\end{document}